\documentclass[twocolumn,twoside,slac_two]{revtex4}
\usepackage{graphicx}
\usepackage{fancyhdr}
\begin{document}

\title{Simulating the Fermi sky: new results after 4 years in orbit}

%

\author{M. Razzano, on behalf of the Fermi-LAT Collaboration}
\affiliation{INFN-Pisa, Largo B. Pontecorvo 3, Pisa, 56127, Italy}
\affiliation{W. W. Hansen Experimental Physics Laboratory, Stanford University, Stanford, CA 94305, USA}

\begin{abstract}
Detailed simulations of the gamma-ray sky played an important role before the launch of Fermi. Pre-launch simulation campaigns aided the development of new analysis tools and the assessment of the expected Fermi-LAT science performance. After four years of the Fermi mission in orbit, improved sky simulations can still be very useful to support the activities of Fermi, to refine our understanding of the scientific performance and to explore potential new science cases. We have used the LAT observations collected so far to develop new, highly-detailed simulations of the sky as seen by LAT. These simulations include the isotropic and Galactic diffuse emission, as well as the list of point and extended sources detected by the LAT. Furthermore, our sky simulations account for transient and variable sources, including Gamma Ray Bursts, pulsars and flaring blazars.  We report on the latest developments of this new sky simulation work, describing its potential benefits for improving the science of Fermi today and in the near future.
\end{abstract}

\maketitle

\thispagestyle{fancy}

\section{INTRODUCTION}
Before the launch of \textit{Fermi} in June 2008, the \textit{Fermi} Large Area Telescope (LAT) Collaboration conducted extensive sky simulation campaigns. Those simulations were very important for the development of the analysis tools and the assessment of the science performance of the LAT~(\cite{atwood2009}). After 4 years of \textit{Fermi} operations, sky simulations are still very valuable to study the science potential of the LAT, because they allow the possibility of performing specific analysis under controlled conditions. The main areas where sky simulations are useful today are:\\
-Catalog-related projects, including studies on sensitivity thresholds or false detection rate;\\
-Instrument-related studies and validations;\\
-Developments and improvements of analysis techniques, such as pulsar blind search algorithms;\\
-Developments of new analysis tools;\\
-Population-related studies (e.g. contribution of faint sources to the diffuse emission);\\
-Evaluation of alternative observing strategies, including dedicated pointed observations;\\
\section{THE LAT OBSERVATION SIMULATOR}
The fundamental tool for creating simulations of the $\gamma$-ray sky as observed by \textit{Fermi}-LAT is the \textit{Observation Simulator} (\texttt{gtobssim}), which is part of the Fermi Science Tools\footnote{http://fermi.gsfc.nasa.gov/ssc/data/analysis/scitools/overview.html}. Gtobssim creates a set of output simulated data products (including photon events and observation intervals), which are formatted into FITS files, exactly as the high-level real LAT data available on the \textit{Fermi Science Support Center} web page \footnote{http://fermi.gsfc.nasa.gov/ssc/}. In order to produce these simulations of detected photons, \texttt{gtobssim} needs three classes of inputs. First of all, it is fundamental to specify a sky model including the $\gamma$-ray sources to be simulated. The details about each simulated source (e.g. sky position, flux, spectrum) are described by XML files, although some sources, like pulsars or Gamma Ray Bursts require additional ASCII files, containing for example the light curve templates. Furthermore, \texttt{gtobssim} needs the LAT Instrument Response Functions (IRFs), which include the Point Spread Function, the Effective Area and the Energy Dispersion. A parameterization of the IRFs as a function of the energy and incidence angle of the incoming photon is thus provided to \texttt{gtobssim} in FITS format. The last piece of information needed by \texttt{gtobssim} is the pointing history of the \textit{Fermi} satellite, which is crucial to determine the portion of the sky observed by the LAT at any time. \texttt{Gtobssim} can accept as input the real pointing history of Fermi in order to create a simulated replica of the real observation history. If no pointing history is provided, \texttt{gtobssim} uses a standard rocking profile and produces a FITS file, which is formatted as the real pointing history of the LAT.\\	
\begin{figure*}[!ht]
\centering
\includegraphics[width=135mm]{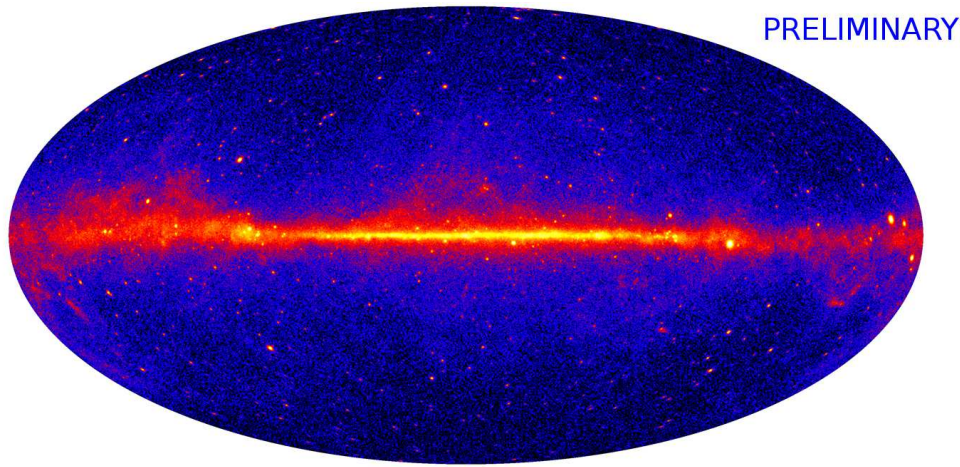}
\caption{\texttt{gtobssim}-based simulation of a 4-year LAT observation of the full sky. The map shows only $\gamma$ rays above 1 GeV that converted in the front part of the LAT tracker. The simulation is based on the real LAT pointing history of the LAT and the P7V6$\_$SOURCE Instrument Response Functions.} \label{fig1_simsky}
\end{figure*}
The distribution of the simulated photons as seen by the LAT is obtained by folding the source models with the IRFs and taking into account the pointing history. The output products, both created in FITS format, are an event file (FT1) containing the list of simulated detected photons with their characteristics (including the reconstructed energy, position, time of arrival).In case no pointing history has been provided by the user, a simulated pointing history file (FT2), based on the standard rocking profile, is also generated.\\
It is important to note that \texttt{gtobssim} does not take into account the detailed propagation of the particles in the LAT (this is done using a tool called Gleam based on the GEANT4 particle interaction codes, as described in~\cite{atwood2009}.

\section{THE SKY SIMULATION INFRASTRUCTURE}
Simulating a large LAT observation of the full sky requires large computational resources, in particular when the number of sources is large and the observation timeline covers few years. Therefore, in order to support massive sky simulations, we developed a dedicated task running on the SLAC Pipeline II framework ~(\cite{atwood2009,zimmer2012}). This task runs parallel jobs, each one simulating a small observation window (e.g. 1 week) and running post-processing operations, including the creation of Good Time Interval (GTIs) and the generation of livetimecube files.\\
We also developed a set of  utilities to manipulate the input sky models or to "post-process" the simulations, such as merging the output files, validating the simulation and creating the relevant documentation.
\begin{figure*}[ht]
\centering
\includegraphics[width=135mm]{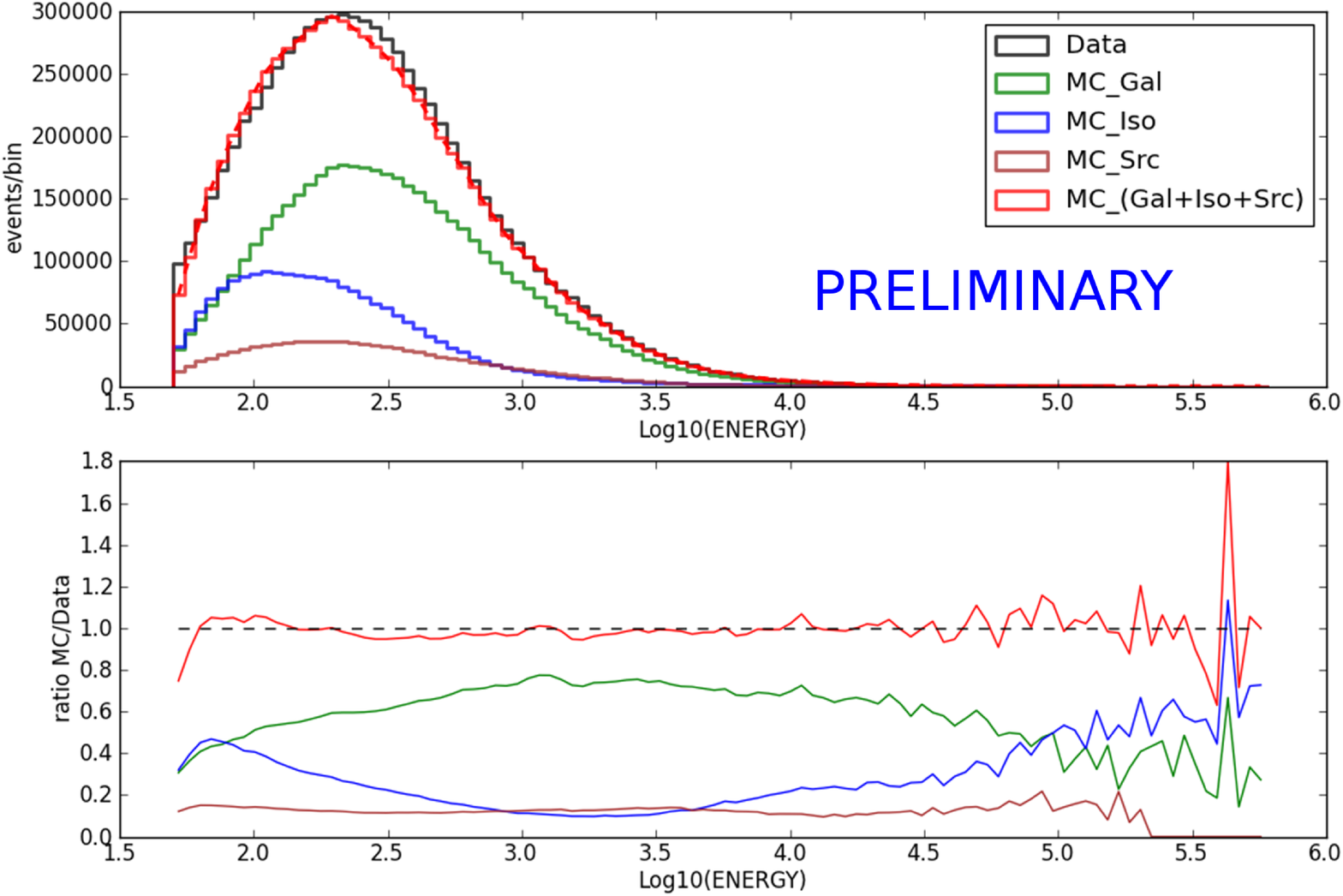}
\caption{Comparison between the energy distributions of real LAT data and \texttt{gtobssim} simulation. This comparison is based on a 6-month simulation. The datasets agree within few percents and the discrepancy is likely due to the absence of the Earth limb emission in the simulations.}
\end{figure*}
\section{SIMULATING THE 4-YEAR SKY OF FERMI-LAT}
We used the infrastructure described in Section 3 to create a realistic simulation of the $\gamma$-ray sky as seen by the LAT during its first four years of operations. This simulation (Fig 1) has the purpose of validating the entire infrastructure and providing a reliable dataset to compare with real data. We included in the sky model many different classes of sources representing the $\gamma$-ray emitters observed by the LAT. In particular, this sky model included the Galactic and the Isotropic diffuse emissions \footnote{http://fermi.gsfc.nasa.gov/ssc/data/access/lat/BackgroundModels.html}, the emission from the quiet Sun\footnote{http://sourceforge.net/projects/stellarics} and the Moon ~(\cite{abdo2011,abdo2012}) and the sources in the Second Fermi Source Catalog (2FGL,\cite{nolan2012}) including the extended sources observed by the LAT (\cite{lande2012}). Furthermore, \texttt{gtobssim} is very flexible and allows the possibility of simulating also variable sources, including Gamma Ray Bursts (\cite{abdo2009}), blazars (\cite{nolan2012}) and $\gamma$-ray pulsars (\cite{razzano2009}). The contribution of unresolved Galactic and extragalactic sources produced by population synthesis codes \footnote{http://www.mpe.mpg.de/~aws/talks/fermi$\_$symp$\_$strong$\_$poster$\_$\\sourcepops.pdf} can also be added (\cite{abdo2010,strong2007}).
Figure 2 shows a comparison between simulated and real data using a dataset of 6 months. The simulation has been created using the P7V6$\_$SOURCE IRFs and the real pointing history of the LAT. The comparison of the distributions of photon characteristics (e.g. the energy) shows that the simulations agree with the data within few percents. This small discrepancy can be due to the absence of the Earth limb model in the simulation and does not affect the overall quality of the sky simulations. We therefore plan to use this sky simulation infrastructure to address more specific studies in the perspective of better understanding the science capabilities of the LAT in the next years of operations.
\bigskip 
\begin{acknowledgments}
The $Fermi$ LAT Collaboration acknowledges support from a number of agencies and institutes for both development and the operation of the LAT as well as scientific data analysis. These include NASA and DOE in the United States, CEA/Irfu and IN2P3/CNRS in France, ASI and INFN in Italy, MEXT, KEK, and JAXA in Japan, and the K.~A.~Wallenberg Foundation, the Swedish Research Council and the National Space Board in Sweden. Additional support from INAF in Italy and CNES in France for science analysis during the operations phase is also gratefully acknowledged.  We would like to thank the DOE SLAC National Accelerator Laboratory Computing Division for their strong support in performing the massive simulations necessary for this work. These sky simulations are the result of a group effort involving many LAT scientists that contribute to the simulation working group, including: Marco Ajello, Jean Ballet, Keith Bechtol, Johan Bregeon, James Chiang, Seth Digel, Nicola Giglietto, Gulli Johannesson, Joshua Lande, Nicola Omodei, Riccardo Rando, Andy Strong, Gino Tosti, Stephan Zimmer.
\end{acknowledgments}

\bigskip 

\end{document}